# Muon spin relaxation and neutron diffraction investigations of quadrupolar and magnetically ordered states of YbRu$_2$Ge$_2$

H. S. Jeevan[1*$], D. T. Adroja[2*], A. D. Hillier[2], Z. Hossain[3], C. Ritter[4] and C. Geibel[1]

[1]Max-Planck Institute for Chemical Physics of Solids, D-01187 Dresden, Germany
[2]ISIS Facility, Rutherford Appleton Laboratory, Chilton, Didcot, Oxon OX11, 0QX, United Kingdom
[3]Department of Physics, Indian Institute of Technology, Kanpur 208016, India
[4] Institut Laue Langevin, BP 156, 38042 Grenoble Cedex 9, France

(Dated: Sept 14$^{th}$, 2011)

The magnetic and quadrupolar ordered states of polycrystalline YbRu$_2$Ge$_2$ have been investigated using zero-field muon spin relaxation (μSR) and neutron diffraction measurements. Specific heat measurements show three successive phase transitions, with decreasing temperature from a paramagnetic to a quadrupolar state at T$_0$ ~ 10 K, from the quadrupolar to a magnetic state at T$_1$ ~ 6.5 K and a possible change in the magnetic ground state at T$_2$ ~ 5.5 K. Clear evidence for the magnetic transition below 7 K (spectrum at 8 K reveals paramagnetic state) and a likely change in the magnetic structure near 5.8 K is observed in the zero-field μSR measurements. The μSR data, however, do not reveal any signature of magnetic order in the temperature range 8 - 45 K. This result is further supported by neutron diffraction measurements, where clear magnetic Bragg peaks have been observed below 8 K, but not above it. Below 8 K, the magnetic Bragg peaks can be characterized by an incommensurate antiferromagnetic ordering with the propagation vector q = [0.352, 0, 0] and the magnetic moment 2.9(3) μ$_B$ of Yb along the b-axis. These results are discussed in terms of quadrupolar ordered and magnetically ordered states.

# I. Introduction

The presence of spin, charge and lattice degrees of freedom in the strongly correlated (f- and d-elctron) systems provide a possibility for ordering mechanisms which are different from the usual magnetic ordering induced through magnetic dipolar interactions [1]. For example quadrupolar [2–4] and octupolar ordering [5, 6] induced through the charge (or orbital) degree of freedom have been observed in several strongly correlated electron systems. Recently these high-rank multipolar ordering of f-electron systems have received increasing consideration in both theoretical and experimental condensed matter physics. In rare-earth intermetallic compounds the interaction between the local moments occurs through conduction electrons by an indirect exchange mechanism, such as the Ruderman-Kittel-Kasuya-Yosida (RKKY) interaction. In most of the systems magnetic dipole ordering is dominating, while in some only quadrupolar ordering (i.e. TmCd) and in a few cases both dipolar and quadrupolar ordering (i.e. TmZn) exist. The occurrence of quadrupolar ordering implies the presence of a quadrupolar moment for the ground state crystal field wave functions. This situation is not very common with only a small number of compounds showing quadrupolar ordering. There are equal numbers of examples of quadrupolar order systems with even number of f-electrons (non Kramers ions) [7-9] and with odd numbers of f-electron (Kramers ions) systems, for example TmTe and the $DyB_2C_2$ compounds [10–11]. In many cases by lowering the temperature systems exhibit first quadrupolar ordering and then by further lowering the temperature they reveal long range magnetic ordering. For example the tetragonal compound $TmAu_2$ (which we will discuss later) shows the ferro-quadrupolar ordering at $T_o = 7$ K and AF magnetic order at $T_N = 3.2$ K [12, 13]. Another example of such behaviour is seen in $DyB_2C_2$, which exhibits AF quadrupolar ordering at $T_Q = 24.7$ K and AF magnetic ordering at $T_N = 15.3$ K [10, 14].

For elements with an odd number of f-electron, like Ce and Yb, the non cubic environment of the f-element splits the ground state J multiples into Kramers doublets. The doublet crystal field ground state of many Ce compounds carries a dipole moment that exhibits a long range magnetic ordering, but a quadrupolar ordering is not common among Ce compounds. A prototype example of quadrupolar ordering in Ce-based compounds is $CeB_6$ which has been extensively studied for many years [2–4]. In contrast, for Yb-based systems no good example for quadrupolar ordering has yet been well established. It is to be noted that YbSb was proposed to exhibit an antiferromagnetic quadrupolar ordering (AFQ) at 5 K [15], but the CEF ground state has no quadrupole moment. Thus the ordering temperature is too high to be of induced type (from the first excited state) where first exited doublet state is above 80 K. A quadrupolar ordering was recently suggested to exist in $YbAl_3C_3$ at 80 K, a temperature which seems extremely high for a quadrupolar transition [16]. Further studies revealed, however, that this transition has a structural origin rather than representing a quadrupolar transition [17, 18]. We recently reported a detailed investigation of the physical properties of single crystalline $YbRu_2Ge_2$ and found that it presents as well anomalous properties [19]. In this compound, Yb is in the 3+ valence state with a quasiquartet CEF ground state. The macroscopic measurements on the single crystal showed an evidence for three phase transitions which were interpreted as reflecting the occurrence of quadrupolar order at $T_0 \sim 10$ K followed by two magnetic phase transitions of presumably antiferromagnetic type below $T_1 \sim 6.5$ K with a possible change in magnetic structure at $T_2 \sim 5.5$ K. Applying an external magnetic field along the a-axis shifts $T_0$ to higher temperatures and broadens the transition while the other two transitions at $T_1$ and $T_2$ shift to lower temperature. The existence of the quasiquartet CEF ground state and the unusual behaviour of $T_0$ in magnetic fields comforted the idea that this transition is related to quadrupolar ordering. Theoretical analysis of $YbRu_2Ge_2$ has shown that ferroquadrupole ordering of either $O_2^2$ or $O_{xy}$ type is possible for the non-magnetic transition at $T_0$ [20]. The authors proposed an incommensurate ordering wave vector from the observed uniform susceptibility in magnetic field parallel to [100] and [110] assuming either ferro $O_2^2$ or ferro $O_{xy}$ order at $T_0$.

In the present work, to probe the nature of these phases, we have performed muon spin relaxation (μSR) [22, 23] and neutron diffraction experiments on a polycrystalline sample of YbRu$_2$Ge$_2$. The μSR and neutron diffraction results, along with other measurements, provide a clear evidence of incommensurate long range magnetic order below 8 K and also support the non-magnetic nature of the ordering at 10 K.

## II. Experimental

The polycrystalline sample was prepared by a solid state sintering method. A stoichiometric (2 % excess of Yb) amount of the three elements (99.99 % pure) was sealed inside a tantalum crucible using arc welding in Ar atmosphere. The sealed Ta crucible was heated and kept at 1200 ºC for four days in Ar atmosphere and then cooled to room temperature. The sample was subsequently powdered and pressed into pellets and heated once more to the same temperature for 6 more days. X-Ray powder diffraction and Energy Dispersive X-ray Analysis (EDAX) were used to check the structure and chemical composition of the sample, confirming the formation of a polycrystalline compound of YbRu$_2$Ge$_2$, with lattice parameter a = 4.2105 (10) Å and c = 9.7567 (20) Å. These values are in good agreement with reported lattice constants from polycrystalline and single crystals [19, 21].

The zero field (ZF) μSR measurements were performed at the pulsed neutron and muon facility at ISIS (Rutherford Appleton Laboratory. UK), using the MuSR spectrometer. An Oxford Instruments Variox cryostat was used to cool the sample down to 1.2 K. The muons are implanted 100% spin polarized with an energy of approximately 3.2 MeV. These muons come to rest rapidly (in $10^{-10}$s) after thermalizing in the sample and precess about the magnetic field at the muon site and subsequently decay into positrons [22, 23]. These positrons are emitted, preferentially, along the spin direction of the muon at the time of decay. The sample was mounted onto a silver plate. Any muons implanted into the silver will give a time and temperature independent signal in the asymmetry. The asymmetry is determined by the equation

$$G_z(t) = (N_F - \alpha N_B) / (N_F + \alpha N_B) \qquad (1)$$

where $N_F$ is the counts in the forward detectors, $N_B$ is the counts in the backward detectors and α is a calibration constant. The neutron diffraction measurements on a polycrystalline sample of YbRu$_2$Ge$_2$ were carried out at various temperatures between 2 and 15 K using the 2-axis diffractometer D20 at the Institut Laue-Langevin in Grenoble, France, using neutrons of wavelength λ= 2.41Å. Rietveld refinements of the powder diffraction data were carried out for the confirmation of the crystallographic structure and the determination of magnetic structures using the program FullProf [24].

## III. Results and discussion

*(a) Specific heat measurements*

Resistivity, magnetic susceptibility and specific heat measurements were carried out to check the presence of all the phase transitions. The temperature dependence of the specific heat ($C_p$) data from 2 K to 15 K of polycrystalline YbRu$_2$Ge$_2$ sample is shown in Fig 1, the inset shows a plot of the entropy below 15 K. The large jump (2 J/mol K$^2$) in the C$_p$/T versus T plot at 6.5 K (T$_1$) marked by an arrow clearly reflects the magnetic order of the Yb$^{3+}$ moments. The magnetic ordering temperature is in good agreement with the transition observed in both polycrystalline and single crystalline magnetic susceptibility measurements. The total entropy at 15 K is ~ 10 J/mol K, which is much above the value expected for a doublet ground state (Rln2 = 5.76 J/mol-K), but close to that expected for a quartet ground state (Rln(4) = 11.52 J/mol-K). It is to be noted that the observed entropy of Rln(2) at 8 K is from the ground state CEF doublets or two CEF-singlet states in molecular field below $T_N$. Strong anomalies are furthermore visible at 10 K ($T_0$) and at 5.5 K($T_2$) respectively (shown by arrows). A detailed discussion of the specific heat data on single crystals has been presented in our previous communication [19].

*(b) µSR measurements*

Figure 2 shows the zero-field µSR spectra at 1.2, 5, 6.5 8, 10 and 45 K, plotted as time dependent muon Asymmetry, along with the fit measured at various temperatures from the polycrystalline $YbRu_2Ge_2$ sample. The spectrum taken in the paramagnetic region at 45 K shows a very weak muon depolarization rate due to fluctuating Yb moments. Even though a slightly stronger depolarisation rate is observed we do not observe any clear sign of frequency oscillations in the asymmetry or a reduction in the initial asymmetry at 10 K or at 8 K (i.e. below $T_0$). The only difference in the µSR spectra is a slight increase in the relaxation rate due to slowing down of the Yb 4f moments at 10 K and 8 K compared to 45 K. This can be taken as further evidence that the transition at 10 K is not due to magnetic order and supports our claim of quadrupolar ordering at 10 K as presented in a previous communication [19]. At 1.2 K a strong depolarization rate with coherent frequency oscillations can be observed thus confirming that the sample is in a long range magnetically ordered state at this temperature. Further the frequency oscillations are also observed at 5 and 6.5 K (Fig.2). All the µSR spectra have been fitted using the following equation

$$G_z(t) = A_1 [\sin(\omega t + \phi)/(\omega t + \phi)] + A_2 \exp(-\lambda_1 t) + C \qquad (2)$$

The first term in Eq.(2) represents the spontaneous muon Larmor precession in the internal field of the ordered state of $YbRu_2Ge_2$. $\omega$ is the muon precession frequency due to local magnetic field acting on the muon, $A_1$ and $A_2$ are the initial asymmetries and $\phi$ is a phase. $\lambda_1$ is the temperature dependent relaxation rate and C is a constant corresponding to the background.

Above 7 K (i.e. just above $T_1$) the zero field muon spectra can be best fitted to $G_z(t)$ with $A_1=0.0$, which implies a single exponential relaxation rate plus a small time independent back ground C. The fit parameters in the paramagnetic phase are shown in Fig. 3 and Fig 4. The inset of Fig 4 clearly shows that at $T_1$, $\lambda_1$ diverges and there is a loss of initial asymmetry. Near and below $T_1$, the μSR spectra require the additional oscillating term. The form of this extra term (first term), given in Eq. (2), has the form of a Bessel function that indicates that the magnetically ordered state is an incommensurate antiferromagnet [23]. $A_1$ (see Fig 3), which represents the magnetic volume fraction convoluted with the instrument frequency response, shows a sharp increase and saturates below 5 K, while on the other hand $A_2$ decreases below $T_1$ due to the loss of paramagnetic volume fraction. Fig 4 shows the precession frequency $\omega$ at different temperatures. The muon precession starts to be measurable at the temperature of the magnetic transition ($T_1$) and increases on lowering the temperature. From the value of $\omega$ it is possible to calculate the internal magnetic field at the muon site and its temperature dependence reflects the development of a magnetic field at the muon site below $T_1$. A clear change in the muon precession frequency near $T_2$ and a peak in the temperature dependence of $\lambda_1$ below 5 K could be associated to a small rearrangement of the spin directions in the magnetic structure (see inset of Fig.4).

**(c) Neutron diffraction study**

In order to investigate the magnetic phase diagram, neutron diffraction patterns were obtained at different temperatures from 15 K (paramagnetic region) down to 1.9 K (much below the three transitions). All diffraction patterns were taken in zero applied magnetic field. The diffraction data at 15 K and 8 K are identical, no new peaks are observed at 8 K. Fig. 5a shows the neutron-diffraction pattern, consisting of only nuclear peaks, of YbRu$_2$Ge$_2$ at 12 K. Refinement of the 12 K spectrum confirms the crystal structure to be tetragonal of ThCr$_2$Si$_2$ type with lattice parameters a = 4.190 (10) Å and c = 9.710 (20) Å slightly smaller compared to the room temperature lattice constant measured by the X-ray diffraction study [20]. There is no sign of a structural transition down to 1.9 K within the resolution of D20. The spectrum measured at 1.9 K and its magnetic part obtained by subtracting the data collected at 12 K is shown in Fig. 5b. All the new reflections which are magnetic in origin can be indexed by a propagation vector q = [0.352, 0, 0], which suggests an incommensurate magnetic order at 1.9 K. The magnetic nature of the reflection can be confirmed as their intensity decreases with increasing wave-vector transfer, Q (or with increasing scattering angle). Magnetic symmetry analysis using the BASIREPS program [24] shows that in the space group I4/mmm and with a magnetic propagation vector of [0.352 0 0] there are only 3 allowed irreducible representations for the Yb atom: The first has a basis vector in x-direction the second in y-direction and the third in z-direction. Refining the neutron data against these different models a very good fit was obtained with the magnetic moment constrained in [010] direction. Assuming q = [0, 0.352, 0] an equivalent fit is achieved with the moment direction in [100] e.g. always perpendicular to the direction of the magnetic propagation vector. The solid line in Fig. 5b shows the fit using the propagation vector q = [0.352, 0, 0] (or equivalent for powder data in a tetragonal system q = [0, 0.352, 0]) with the tick marks showing the positions of the expected magnetic Bragg peaks. The bottom line shows the difference between the experimental and calculated patterns. Mixing different irreducible representations by allowing the magnetic moment to have components along the x- or z-direction did not improve the refinement and returned as expected on the basis of the symmetry analysis zero moment values for these directions. The sin-wave type of magnetic structure sees the Yb-moments oscillating between zero and a maximum value of 2.9(3) $\mu_B$. This value of the magnetic moment is smaller than the ordered moment of the $Yb^{3+}$ ion $g_JJ$ = 4 $\mu_B$. The observed small value of the magnetic moment of the Yb ion is due to the presence of crystalline electric field that has been seen in our inelastic neutron scattering (INS) study [25]. Our INS study carried out on the HET spectrometer reveals a strong CEF excitation near 31.7 meV. The magnetic structure of $YbRu_2Ge_2$ is shown in Fig.6, which reveals the spin-density wave modulation of the magnetic moment of the Yb atom along the a-axis (modulation direction, but the moment is along b-axis). The magnetic Bragg R-factor was found to be 6.82% from the fit of the magnetic peaks only. All the magnetic Bragg peaks disappear at 7.8 K, and even after collecting data for a longer time at 8 K, we could not find any evidence of magnetic Bragg peaks or new structural peaks at 8 K. This establishes that, within the sensitivity of both neutron diffraction and µSR, the phase transition at 10 K can not be associated with magnetic order. Fig. 7(a) shows the intensity of the strongest magnetic peak as a function of temperature. Interestingly, no discontinuous change in the temperature dependence of the intensity is observed at $T_2$, however, the peak position of this magnetic Bragg peak shows a small shift (Fig.7(b)) indicating a small change in the value of the propagation vector at this temperature. The temperature dependence of the propagation vector estimated form the Rietveld refinement of the data at selected temperature points is shown in Fig. 8, which reveals the change in the slope between 5.1 and 5.9 K. This is in good agreement with the change in the precession frequencies of muons in this temperature range. To determine the small change in the propagation vector of the magnetic structure neutron diffraction measurements on single crystal YbRu$_2$Ge$_2$ are highly desirable.

Here we compare the direction and the value of the moment (b-axis) estimated from our neutron diffraction data with the moment direction predicted in Takimoto and Thalmeier model for pure CEF ground state [20]. The pure CEF $\Gamma_6 = |7/2, \pm1/2>$ taken by ref. [20] as close approximation to the real CEF ground state where some admixture of the $|7/2, \pm7/2>$ is expected, bears a saturation moment of 0.6$\mu_B$ along the c direction and 2.3$\mu_B$ in the basal plane. The latter value is smaller than the value we obtained from our fit of the neutron data. However, because the first excited CEF levels is at a very small energy, the molecular field shall result in this first excited CEF level being involved in the formation of the staggered AFM moment, which should result in a larger value of the ordered moment. This effect can be estimated by analyzing the magnetization for H //[100] at 4 K calculated in ref. [20]. At an applied magnetic field of 5 T the magnetization is bending towards a saturation value of 3.5$\mu_B$, which is likely close to the value of the ordered AFM moment. Thus moment value and the direction predicted in Takimoto and Thalmeier model [20] is in agreement with that estimated with our neutron diffraction study. It is important to note that the added quadrupolar order in the model introduces exchange anisotropy for planar magnetic moments. Finally we would like to mention here that our preliminary analysis of the inelastic neutron scattering data at 8 K also gives the ground state moment of $<\mu_x>$=3.2 $\mu_B$ with ground state wave functions dominated by $|\pm1/2>$ with small component of $|\pm7/2>$.

**IV. Conclusions**

In conclusion, we have carried out μSR and neutron diffraction measurements on YbRu$_2$Ge$_2$ to investigate the nature of three phase transitions observed in the specific heat at 10 K, 6.5 K and 5.5 K. Both μSR and neutron diffraction studies confirm that the phase transition at 10 K is non-magnetic in origin. This is in agreement with our previous proposed scenario of a quadrupolar nature of the transition. Within the resolution of our present diffraction study we have not seen any sign of a structural change below 10 K. μSR and neutron diffraction studies confirm and describe the long range magnetic ground state below 8 K. The neutron data suggest that the weak peak seen at 5.5 K in the temperature dependence of the μSR frequency may reflects a possible small change in either the propagation vector of the sin-wave modulated magnetic structure or a very small change in the spin direction. We have also compared the value of the observed Yb moment form our neutron diffraction data with that estimated from the theoretical model in ref. [20] and found good agreement. It is interesting to point out that for a stable local moment an amplitude modulated structure cannot be retained down to T = 0 K, for entropy reason it has to evolve e.g. to a square up structure [26]. In the present case, on the one hand the Kondo interaction by partially screening the 4f moments makes an amplitude modulated structure possible down to T = 1.9 K in YbRu$_2$Ge$_2$ as seen through our diffraction study. On the other hand the high order peaks connected with a squared up structure are likely too weak to be observable in the present powder diffraction patterns or a possibility of squared structure below 1.9 K. Further experiments are planned in order to elucidate the nature of the 10 K transition.

**Acknowledgement**

We would like to thank to Drs W Kockelmann, A Daoud-Aladine, Gudrun Auffermann and Prof. L Chapon for their assistance in the neutron diffraction analysis and Dr M. Rotter for an interesting discussion on crystal field. Financial assistance from BRNS and NMI3 is also acknowledged.

$ Present address: Physikalisches Institut, Georg-August-Universität Göttingen, D-37077 Göttingen, Germany

**Figure Captions**

Fig. 1. Temperature dependence of the specific heat of polycrystalline YbRu$_2$Ge$_2$, plotted as $C_p/T$ versus T showing the presence of three transitions. Arrows indicate the position of the phase transitions as described in the text. The inset shows a plot of the entropy versus temperature.

Fig. 2. (color online) µSR spectra of YbRu$_2$Ge$_2$ collected at four different temperatures 45 K (squares), 10 K (diamonds), 8 K (down triangles), 6.5 K (hexagons), 5 K (up triangles) and 1.2 K (circles), showing a very small depolarization rate at 45 K, 10 K and 8 K and a strong depolarization with coherent frequency oscillations between 1.2 K and 5 K. The lines are fit to the data using Eq. (2).

Fig. 3. (color online) Temperature dependence of the initial muon asymmetries of the two components of Eq. (2). The initial asymmetry amplitudes are associated with the paramagnetic $A_2$ and magnetic $A_1$ volume fractions. A loss of asymmetry $A_2$ is observed at $T_1$.

Fig. 4. Temperature dependence of the muon frequency (MHz). The anomaly observed at $T_2$ ~5.8 K indicates a small change in the magnetic structure. The inset shows the temperature dependence of relaxation rate $\lambda_1$.

Fig. 5. (color online) Neutron-diffraction pattern of $YbRu_2Ge_2$ (a) recorded at 12 K, above the magnetic phase transition (points) together with the best fit (the full line), and (b) difference spectrum 1.9 K - 12 K showing the purely magnetic peaks. The new reflections of magnetic origin can be indexed with the propagation vector q = [0.352 0 0]. The full line at the bottom represents the difference between the experimental data and the fit. Please note that in (b) the difference of 1.9 K-12 K had a negative base line, which gave a problem in Fullprof program, hence the data were shifted by 6000. The negative difference is due to the fact that the background at 1.9 K (in the magnetic ordered state) is lower than that of at 12 K (in the paramagnetic state). This is due to the transfer of the paramagnetic intensity in the magnetic Bragg peaks below the magnetic ordering. Also the peak near $2\theta$~23º is due to the imperfect subtraction of the impurity peaks (due to change in the back ground).

Fig. 6. (color online) The spin-density wave type magnetic structure of $YbRu_2Ge_2$ at 1.9 K. For clarity only Yb atoms have been shown. The Yb-moments are along the b-axis and they modulate along the a-axis with a propagation vector q = [0.352 0 0]. Please note that the magnetic structure is the transverse spin density wave.

Fig. 7. (color online) (top) The integrated intensity of the magnetic Bragg peak indexed as [0.352 0 0] as a function of temperature. The solid symbols are data collected with a long counting time, while open symbols are data collected for a shorter counting time. (bottom) The temperature dependence of the [0.352 0 0] peak. The vertical dotted line is guide to eyes.

Fig. 8. The temperature dependence of the propagation vector obtained from the Rietveld refinement of the diffraction data at selected temperature points. The dotted lines are guide to eyes.

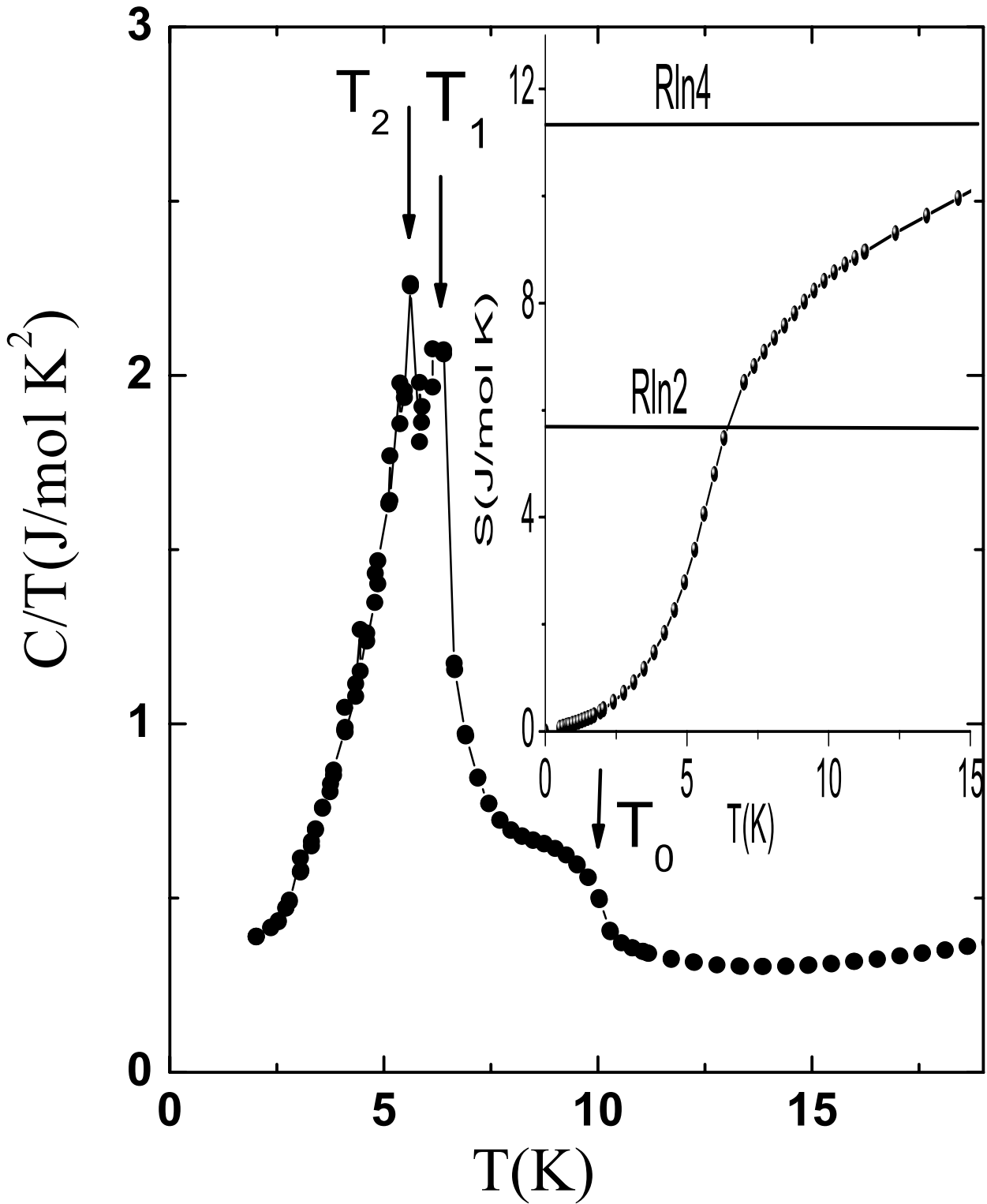

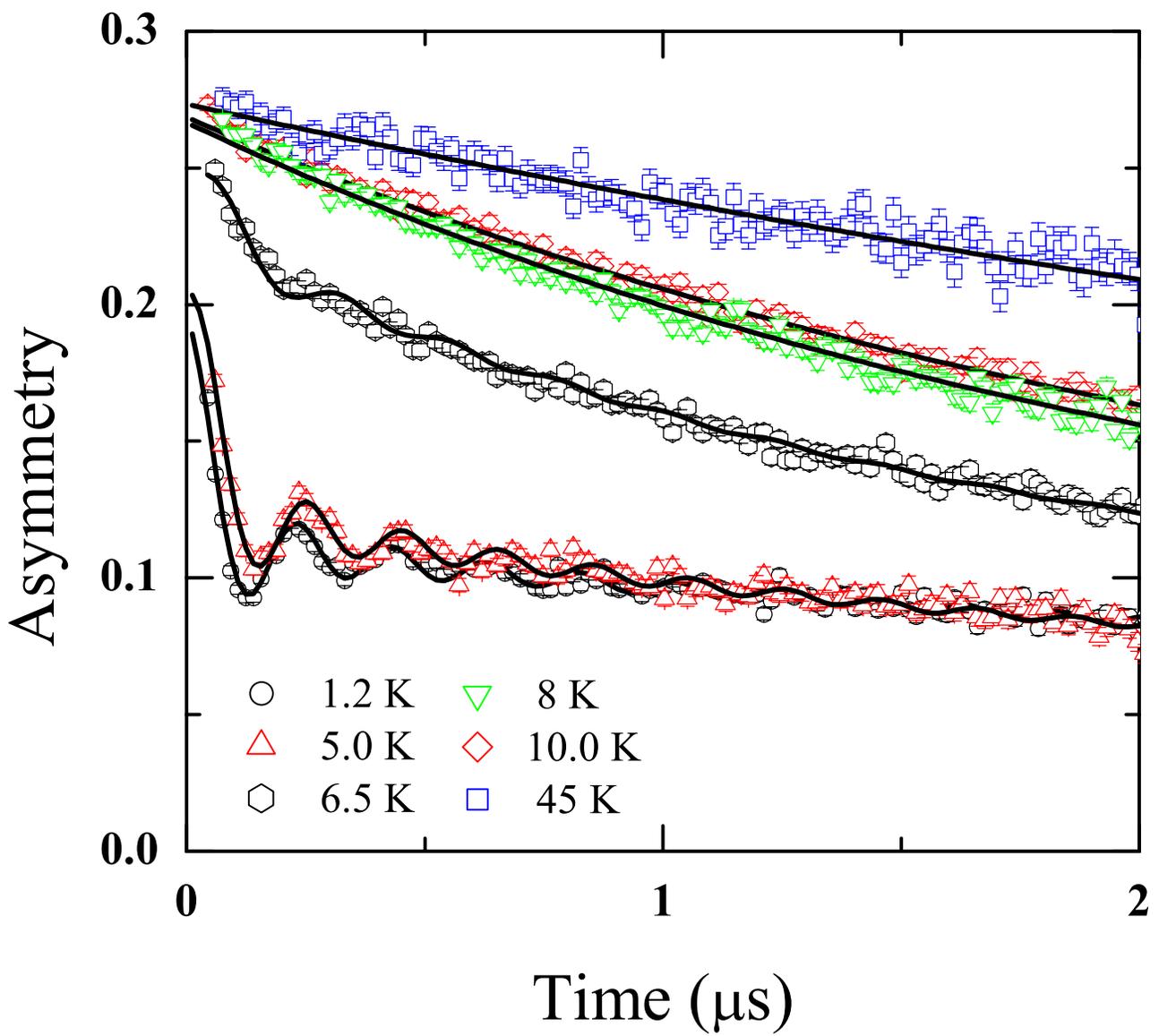

Fig.3 Jeevan et al

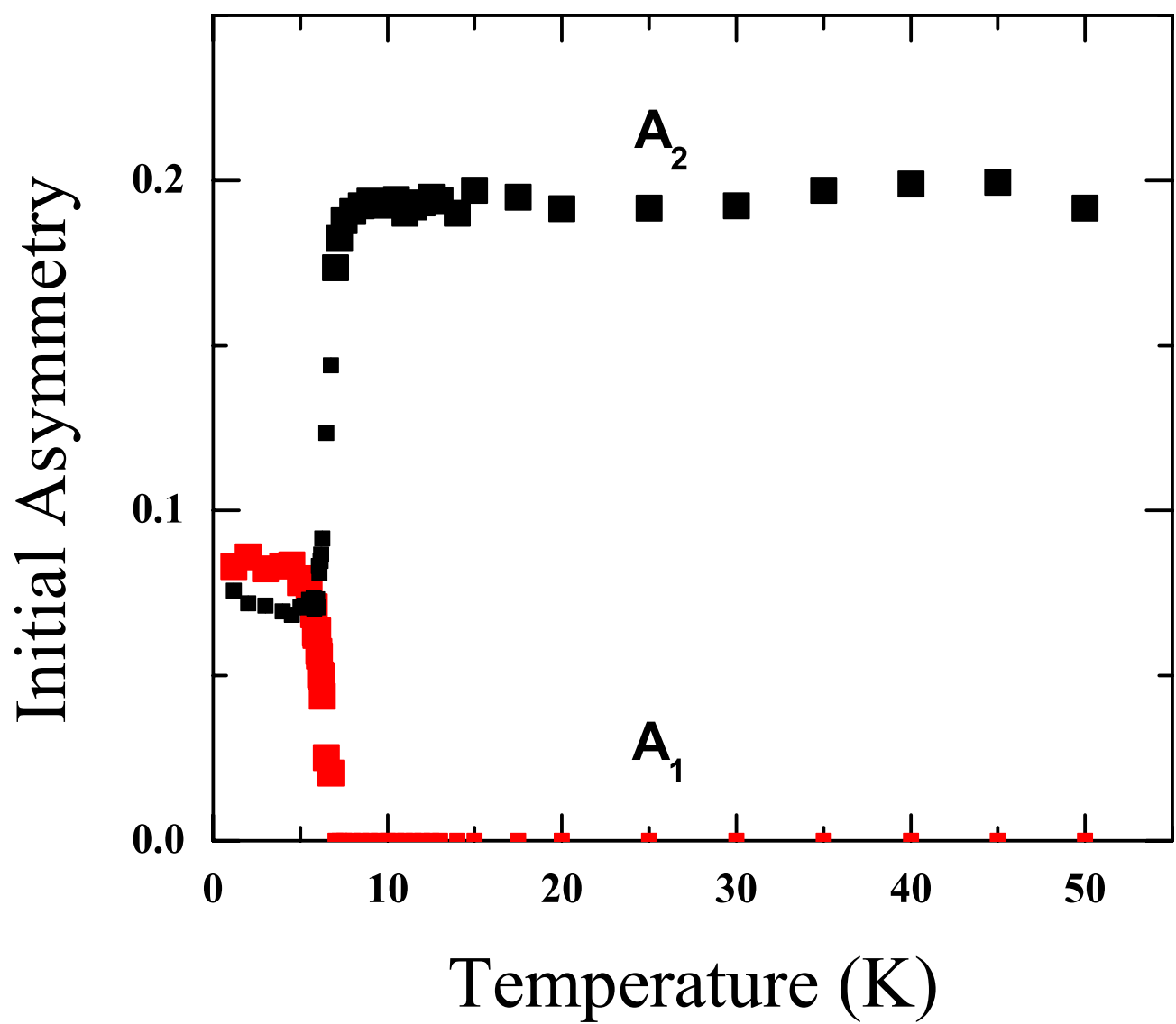

Fig.4 Jeevan et al

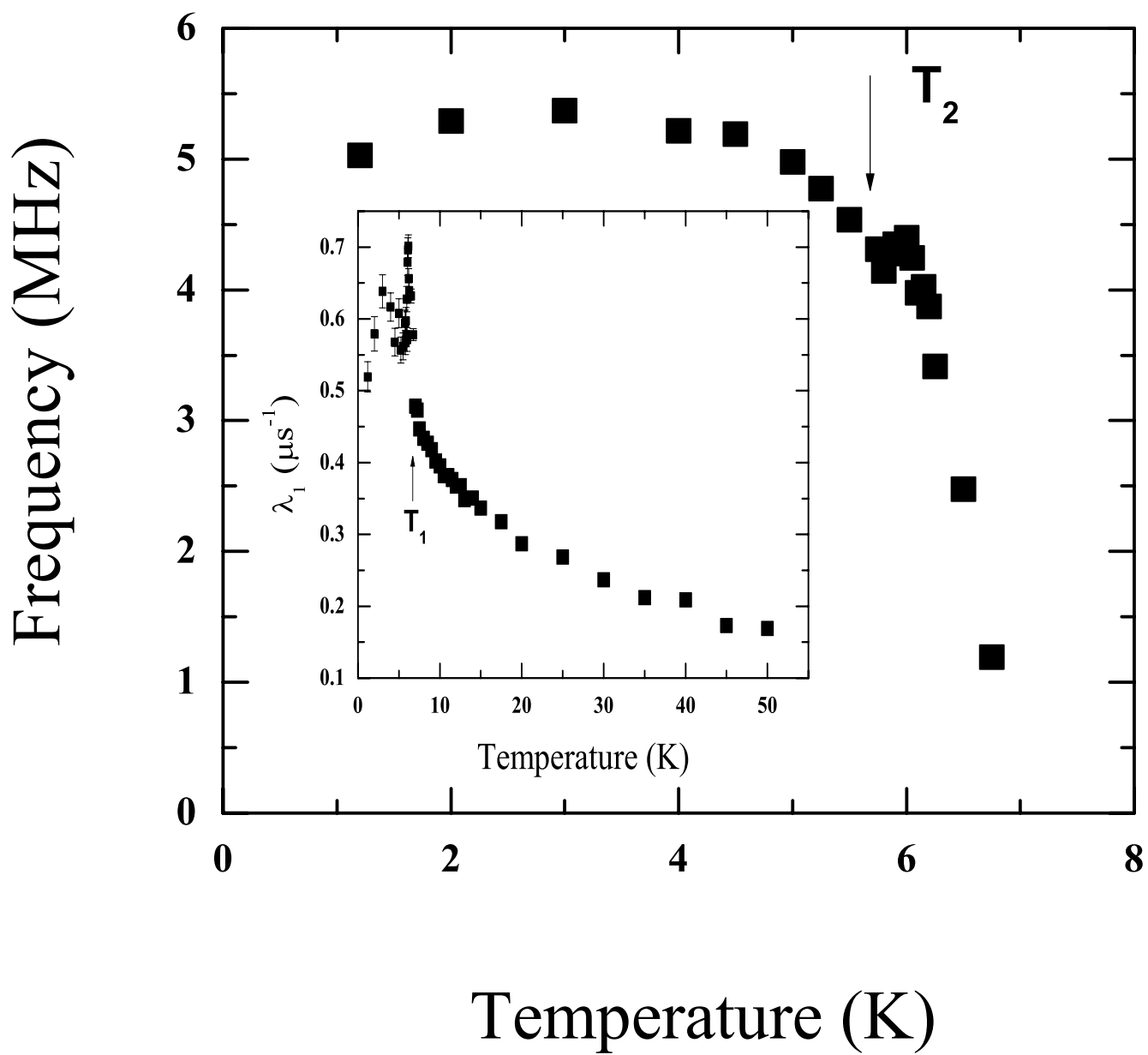

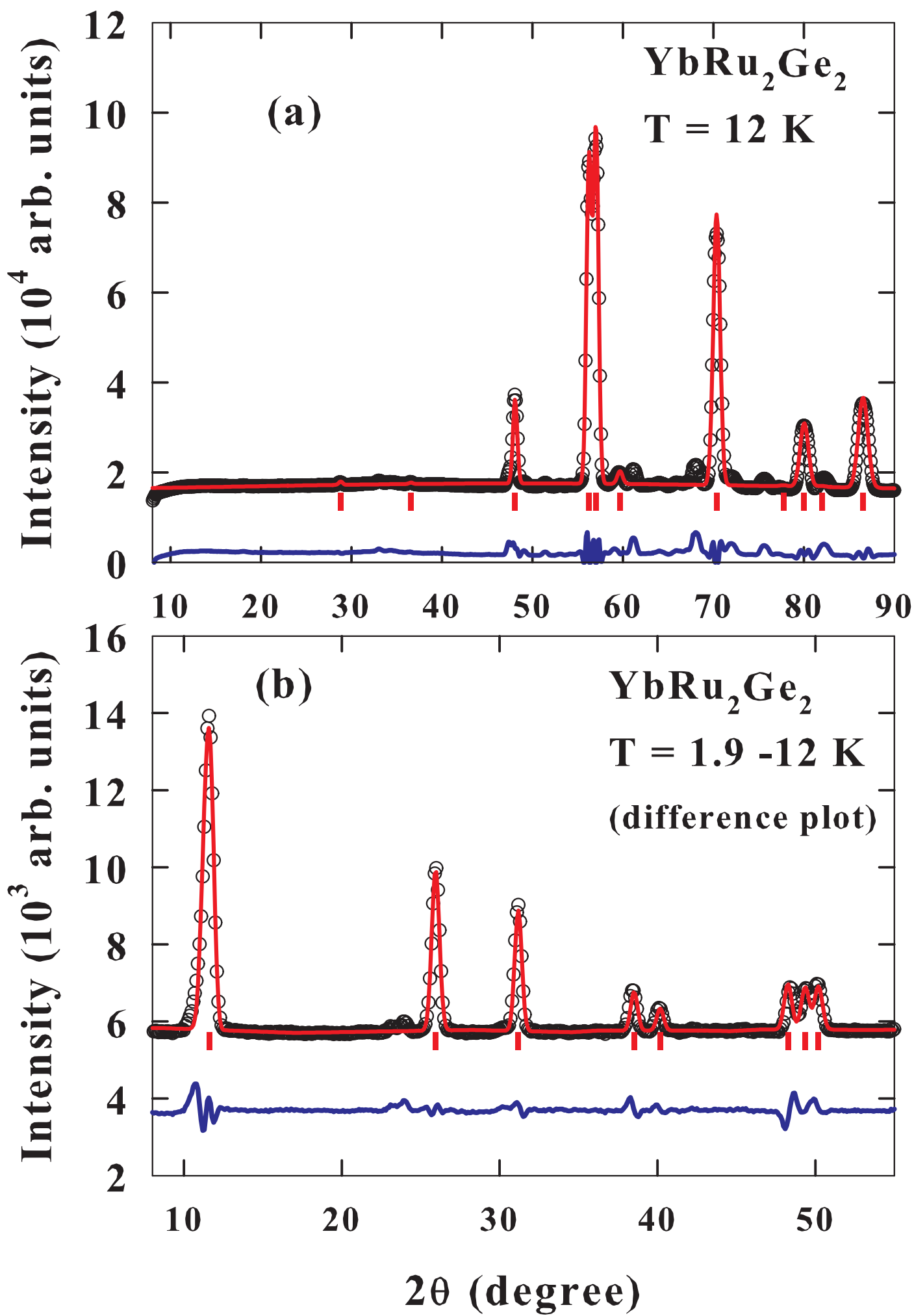

Fig. 6 Jeevan et al

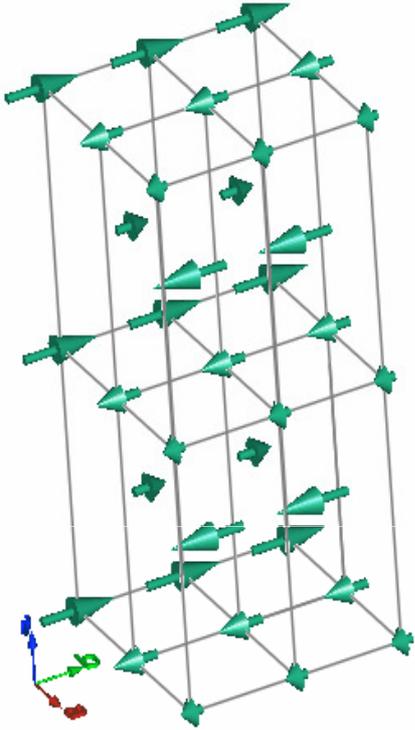

Fig. 7 Jeevan et al

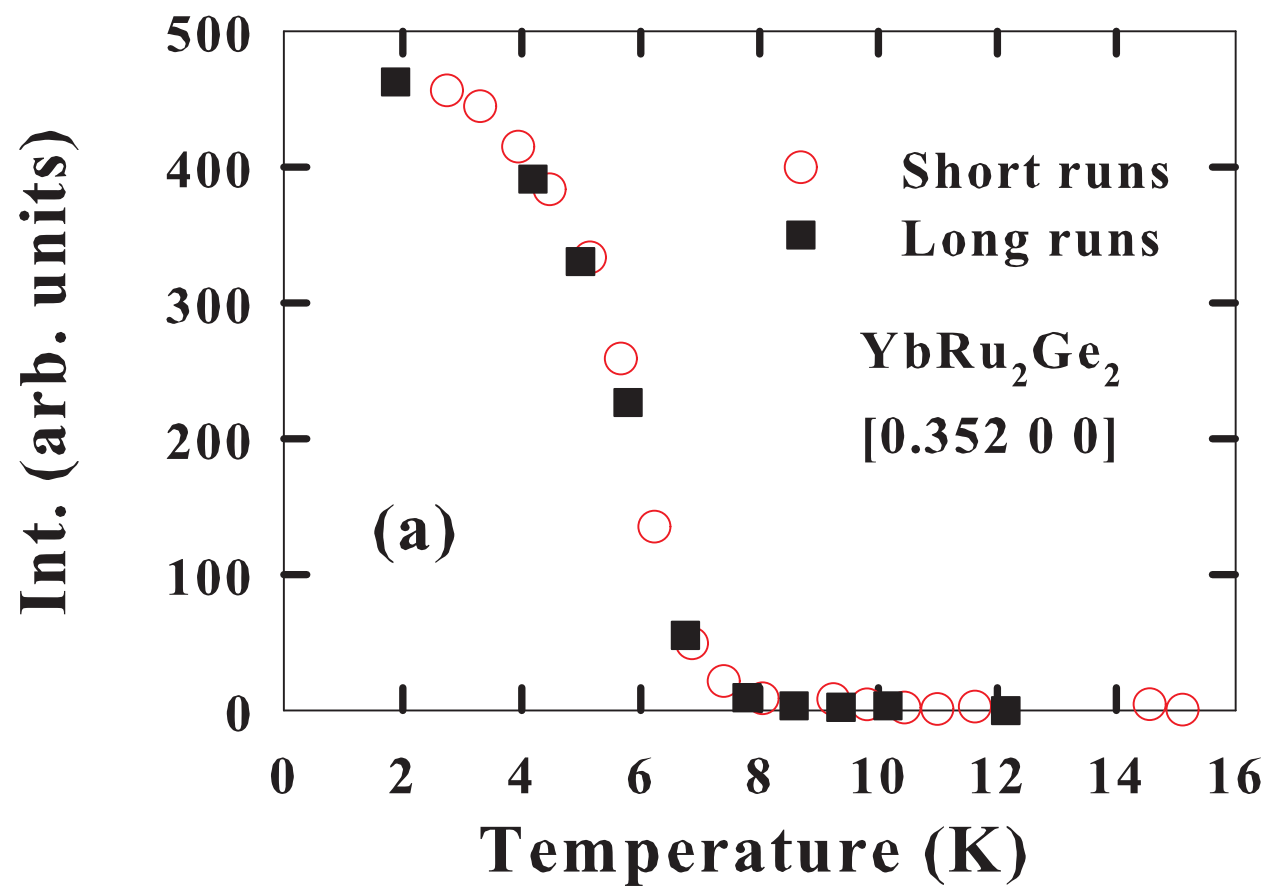

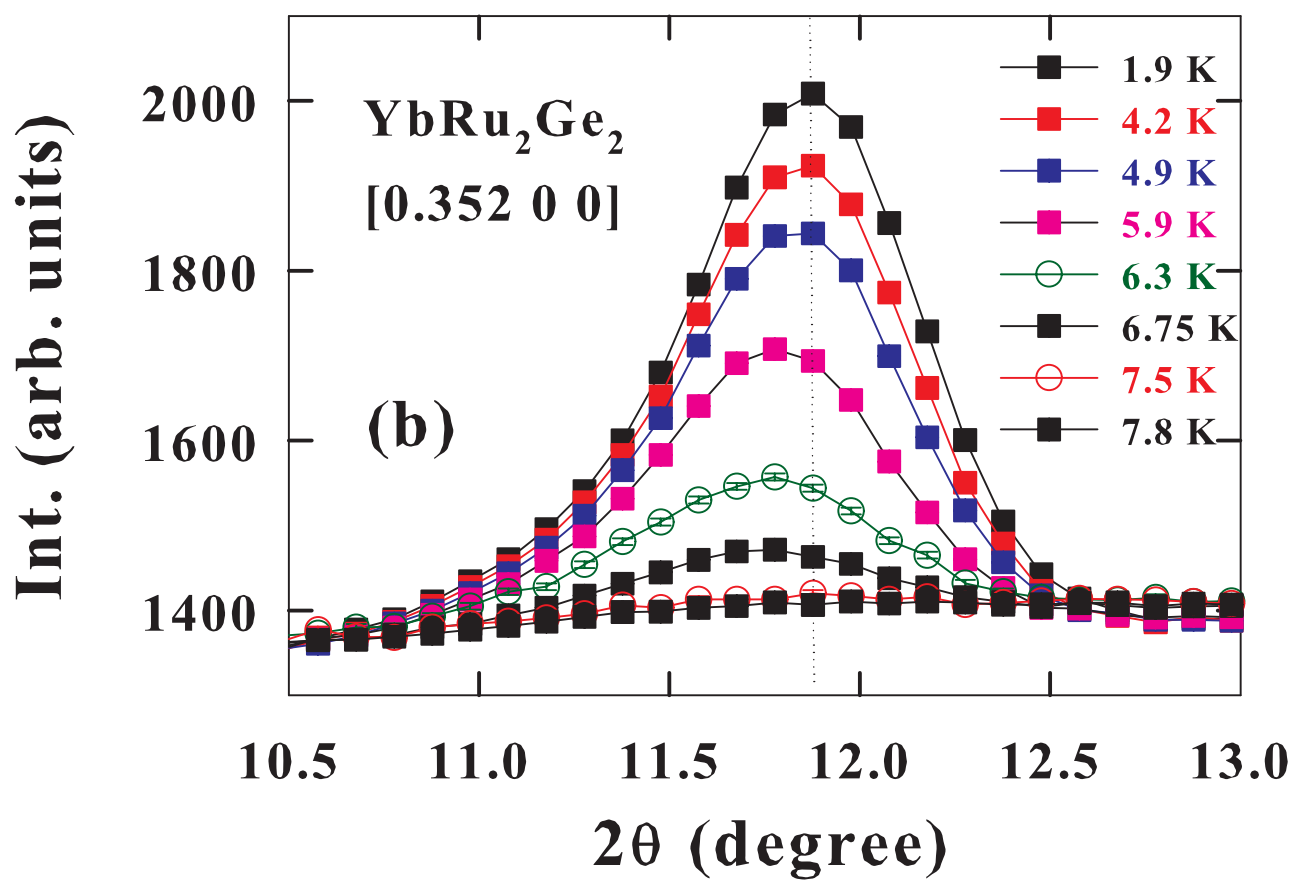

Fig.8 Jeevan et al

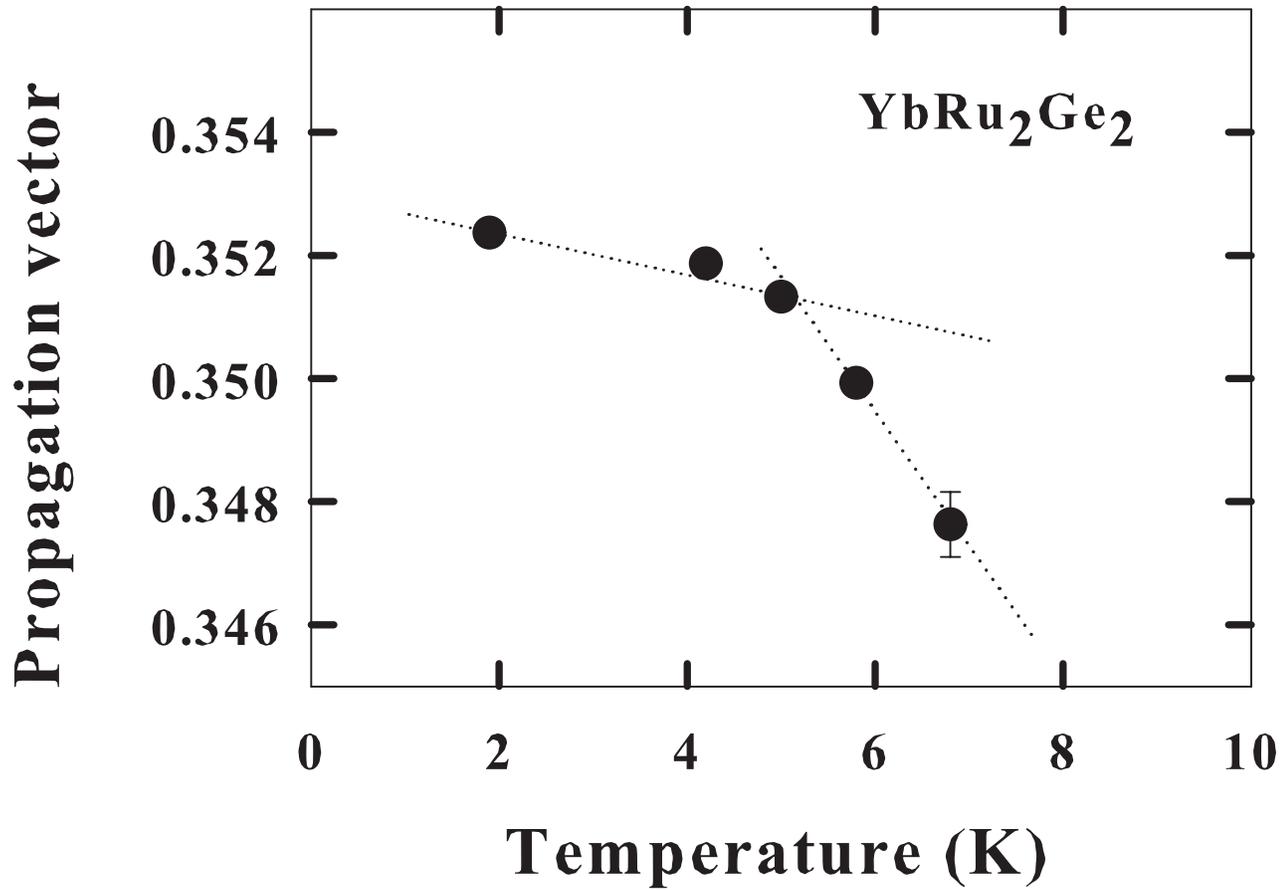